\documentclass[conference]{IEEEtran}
\IEEEoverridecommandlockouts
\renewcommand{\baselinestretch}{0.99}
\usepackage{cite}
\usepackage{amsmath,amssymb,amsfonts}
\usepackage{algorithm}
\usepackage{algpseudocode}
\usepackage{graphicx}
\usepackage{textcomp}
\usepackage{xcolor}
\usepackage{comment}
\usepackage{csquotes}
\usepackage{url}
\usepackage{hyperref}
\def\BibTeX{{\rm B\kern-.05em{\sc i\kern-.025em b}\kern-.08em
    T\kern-.1667em\lower.7ex\hbox{E}\kern-.125emX}}
\begin{document}


\title{IoT-Enabled Smart Car Parking System through Integrated Sensors and Mobile Applications}

\author{\IEEEauthorblockN{Abdullah Al Mamun\IEEEauthorrefmark{1}, Abdul Hasib\IEEEauthorrefmark{2}, Abu Salyh Muhammad Mussa\IEEEauthorrefmark{3}, Rakib Hossen\IEEEauthorrefmark{4}, and Anichur Rahman\IEEEauthorrefmark{5}}

\IEEEauthorblockA{ \textit{Dept. of Internet of Things and Robotics Engineering, Bangabandhu Sheikh Mujibur Rahman Digital University, Bangladesh,}\\
\textit{Dept. of Cyber Security Engineering, Bangabandhu Sheikh Mujibur Rahman Digital University, Bangladesh,} \\
\textit{Dept. of Computer Science and Engineering, National Institute of Textile Engineering and Research (NITER)}, \\
\textit{Constituent Institute of Dhaka University, Savar, Dhaka-1350, Bangladesh}
\\
shaikat.bdu@gmail.com\IEEEauthorrefmark{1}, 
sm.abdulhasib.bd@gmail.com\IEEEauthorrefmark{2},  
abusalyh43@gmail.com\IEEEauthorrefmark{3},
rakib0001@bdu.ac.bd\IEEEauthorrefmark{4},
anis\_cse@niter.edu.bd\IEEEauthorrefmark{5}}}   

\maketitle

\begin{abstract}
Due to more population congestion and car ownership, the provision of parking spaces for vehicles is becoming a crucial factor. This paper aims to present a novel Internet of Things (IoT)--based smart car parking system that can effectively manage these problems with the help of sensor technology and automation. Infrared (IR) sensors,  DHT22 sensors, MQ-2 gas sensors, and servo motors are used in the parking space.  An OLED display shows the status of parking slots in real-time. Communicating with a mobile application through the Message Queuing Telemetry Transport (MQTT) protocol enables the efficient exchange of data. As a result, this innovative solution optimizes parking space, increases efficiency, and makes the parking lot more comfortable. This IoT system allows real-time monitoring and automation of parked cars as well as fast response to dynamic changes in environmental conditions, setting a new standard for smart parking systems.

\end{abstract}

\vspace{2mm}

\section{Introduction}
The majority of people on the planet reside in cities. Therefore, cities are now fully occupied. According to our research on transport and environment, passenger car transit increased by 18\% from 2000 to 2019 and road freight by 31\% \cite{r1}. The global smart parking market is expected to be valued at around US\$5.1 billion in revenue in 2023. According to IMARC Group, the market is projected to grow to nearly \$13.8 billion by the close of 2032, at a CAGR of 11.3\% during the forecast period (2024–2032) \cite{r2}. Cities are currently responsible for more than 75\% of the garbage produced, 80\% of CO2 emissions, and 75\% of the energy consumed \cite{r3}.  According to EEA data, CO2 emissions from passenger automobiles in the 27 EU Member States increased by 5.8\%, while emissions from heavy freight vehicles climbed by 5.5\%, from 2000 to 2019  \cite{wadud2024garduino, r4}. According to researchers, traffic congestion caused by persons looking for parking accounts for 30 percent of road transport emissions. Previous research reveals that hunting for parking produces 30\% of the congestion in populated places. Drivers spend an average of 7.8 minutes looking for a spot \cite{r5}. 
Moreover, in the parking lots of shopping complexes, many particularly unkind drivers who are agitated by the inadequate availability of parking spaces frequently occupy the parking spaces specifically made available for physically challenged people such as the aged and disabled, hence encouraging anti-social behavior. 
Smart parking systems can increase the quality of users’ lives by simplifying their daily activities, reducing fuel consumption and emissions, and mitigating congestion in cities. These parking systems will help ease traffic congestion by providing information on the availability of parking spaces in public and private facilities, thus giving comfort to users and improving user experience \cite{r8, rahman2024blocksd}. 

\par On the other hand, IoT is a key enabler in realizing the concept of smart cities, where one of its important components is enhancing car parking and traffic facilities \cite{r9}. It has become a challenge for drivers in modern-day cities to look for a parking slot due to the excessive amount of private car users. Smart cities are supposed to rationalize their parking assets, which will help minimize searching, congestion, and accidents as well. If drivers are provided with prior knowledge of available parking slots at their desired location and nearby destinations, this will ultimately fix the problems related to parking and traffic congestion \cite{islam2021blockchain, rahman2021smartblock}. 
Indeed, smart parking solutions play a critical role in achieving enhanced urban mobility and a sustainable environment.\vspace{1mm}

In this paper, an IoT-enabled smart car parking system through integrated sensors and mobile applications is proposed. This system includes IR sensors for the exact detection of vehicles, DHT22 sensors for temperature and humidity, and MQ2 gas sensors to detect hazardous gases. Core components of Arduino UNO and Raspberry Pi microcontrollers are responsible for gate operations, ventilation, and data communication via MQTT. The mobile app will provide real-time parking availability information to users, improving the overall user experience. In this regard, the main contributions of the paper could be understood as follows:

\begin{itemize}
    \item In this paper, IR sensors are employed for vehicle detection, DHT22 sensors for temperature and humidity measurement, and MQ2 sensors for hazardous gases to ensure real-time data. \vspace{1mm}

    \item The authors make use of the Arduino UNO and Raspberry Pi microcontrollers in automating gate operations with exhaust fan management at the center of the Smart Car Parking System to monitor environmental conditions.\vspace{1mm}

    \item We use the MQTT protocol in this Smart Car Parking System to provide real-time updates through a mobile app, which shows parking availability and environmental conditions on an OLED screen.\vspace{1mm}

    \item This system demonstrates high accuracy in vehicle detection and environmental monitoring, significantly improving parking management efficiency and user satisfaction with automated, real-time updates.

\end{itemize}
Overall, the proposed solution combines automation, IoT technology, and real-time data processing for improved operational performance and user convenience. It provides a stable and scalable response to contemporary parking difficulties.

\vspace{1mm}
\textbf{Organization:}  This paper is structured as follows: Section II  provides an in-depth review of related work.
Section III presents the methodology, system architecture, and detailed implementation of our IoT-based smart parking solution. Section IV discusses the real-time analysis and experimental results. 
Further, Section V addresses the limitations encountered during the study and proposes potential future enhancements. Finally, this paper concludes in Section VI.

\begin{table*}
\centering
\tiny
\scriptsize
\caption{Comparison with the Existing Smart Parking Systems}
\label{tab:comparison}
\begin{tabular}{|p{1.7cm}|p{2.4cm}|p{2.4cm}|p{2.3cm}|p{2.5cm}|}
\hline
\textbf{Study} & \textbf{Technologies Used} & \textbf{Key Features} & \textbf{Strengths} & \textbf{Weaknesses} \\
\hline
Nikitha et al. (2021) \cite{l2} & Raspberry Pi, IR sensors, camera & Vehicle detection, license plate capture & High accuracy & No online booking, limited sensors \\
\hline
Won et al. (2021) \cite{l8} & IR sensors, cloud, mobile app & Real-time updates, remote reservation & Reduces search time, fuel & Sensor calibration, internet dependency \\
\hline
Sofian et al. (2022) \cite{l3} & ESP32, light sensors (LDR) & Space detection, web app updates & Effective in crowded areas & Complex wiring, sensor sensitivity \\
\hline
Simatupang et al. (2022) \cite{l9} & ESP32, ultrasonic sensors & Online slot check, reservation & Enables reservations & Update delay, unauthorized parking \\
\hline
Venu et al. (2022) \cite{l10} & RFID, IR Sensors, Node MCU, Firebase, MIT App Inventor & Slot monitoring, remote booking, RFID access, cloud sync & Efficient, secure, remote access, real-time updates & Potential sensor errors, setup costs \\
\hline
Vinayak et al. (2022) \cite{l7} & RFID, Raspberry Pi & Space identification, vehicle tracking & Centralized tracking & High setup costs, maintenance \\
\hline
Amran et al. (2023) \cite{l1} & IR sensors, ESP32, servo motors & Space availability, gate control & Improved user experience & Technical reliability, user adherence, scalability \\
\hline
Singh et al. (2023) \cite{l11} & Arduino, ESP8266, IR sensors & Real-time mapping, slot updates, remote booking & Less intervention, efficient space use, user-friendly, real-time info & Sensor accuracy, connectivity issues, setup costs, RFID misuse \\
\hline
Kusuma et al. (2023) \cite{l5} & Arduino, ESP32-CAM, ultrasonic, PIR sensors & Touchless operation, gate control & High success rates & WiFi dependency, sensor angle limitations \\
\hline
Shaikh et al. (2024) \cite{l6} & Raspberry Pi, Pi camera, ultrasonic sensor & Real-time detection, data processing, occupancy verification & Reduces search time, real-time updates & Sensor calibration, setup complexity \\
\hline
Proposed system & IR sensors, DHT22, MQ-2, Raspberry Pi, mobile app, servo motors, buzzers, OLED, MQTT & Vehicle detection, environment monitoring, automated control, spot availability indication & Holistic approach, real-time data, seamless integration & Setup complexity, MQTT dependency \\
\hline
\end{tabular}
\end{table*}

\section{Related Works}
Various systems were proposed to address smart-parking problems. Nikitha et al. \cite{l2} developed a Raspberry Pi-based system using IR sensors and an LCD display without online booking and advanced sensor integration. Won et al. \cite{l8} introduced an IoT system implemented using IR sensors, cloud databases, and a mobile app providing real-time updates and remote reservations but with a dependency on sensor calibration and internet stability. Meanwhile, Sofian et al. \cite{l3} introduced the ESP32-based system that uses light sensors for real-time parking data on a website and a mobile app, yet faced the problem of overwhelming assembling and sensor sensitivity. Simatupang et al. \cite{l9} designed a system with ESP32 and ultrasonic sensors for online reservations, but update delays and wrongful parking were the issues. In the same vein,  Venu et al. \cite{l10} designed RFID and IoT-based secure entry as well as cloud-based booking, which faced the challenges of sensor failure and low-internet connectivity. Vinayak et al.\cite{l7} developed RFID and Raspberry Pi implemented solutions facing high setup costs as well as heavy maintenance problems.

Other works are also notable like Amran et al.'s \cite{l1} system that employed infrared sensors, ESP32 microcontrollers, and servo motors for parking space availability, but the system was found not reliable and not scalable. Singh et al. \cite{l11} presented SCPS with IR sensors, Arduino, and NodeMCU ESP8266, which offered real-time booking but were difficult to maintain due to accurate sensors required both for booking as well as connectivity. Kusuma et al. \cite{l5} developed a touchless system using Arduino and ESP32-CAM, but it suffered from a lack of WiFi stability as well as sensor angles. Shaikh et al. \cite{l6} proposed a system using Raspberry Pi, camera, and ultrasonic sensor for reducing parking search time, but again, the system had the disadvantage of WiFi dependency and the need for calibration.\vspace{2mm}

The proposed IoT-enabled smart car parking system uses advanced features like IR sensors for vehicle detection, DHT22 sensors for temperature and humidity, and MQ-2 sensors for air quality. It uses servo motors for automated gate control, buzzers for entry alerts, and an OLED display for real-time slot availability. The MQTT-connected mobile app allows real-time monitoring and control, outperforming traditional non-IoT solutions in integration, automation, and responsiveness. Table \ref{tab:comparison} highlights the differences and similarities between our work and other studies, showing how our system integrates various sensors and communication technologies to enhance the parking system \cite{rahman2022sdn}.

\section{Proposed Methodology for IoT-Enabled Smart Car Parking System}

\begin{figure*}[h]
\centerline{\includegraphics[scale=0.35]{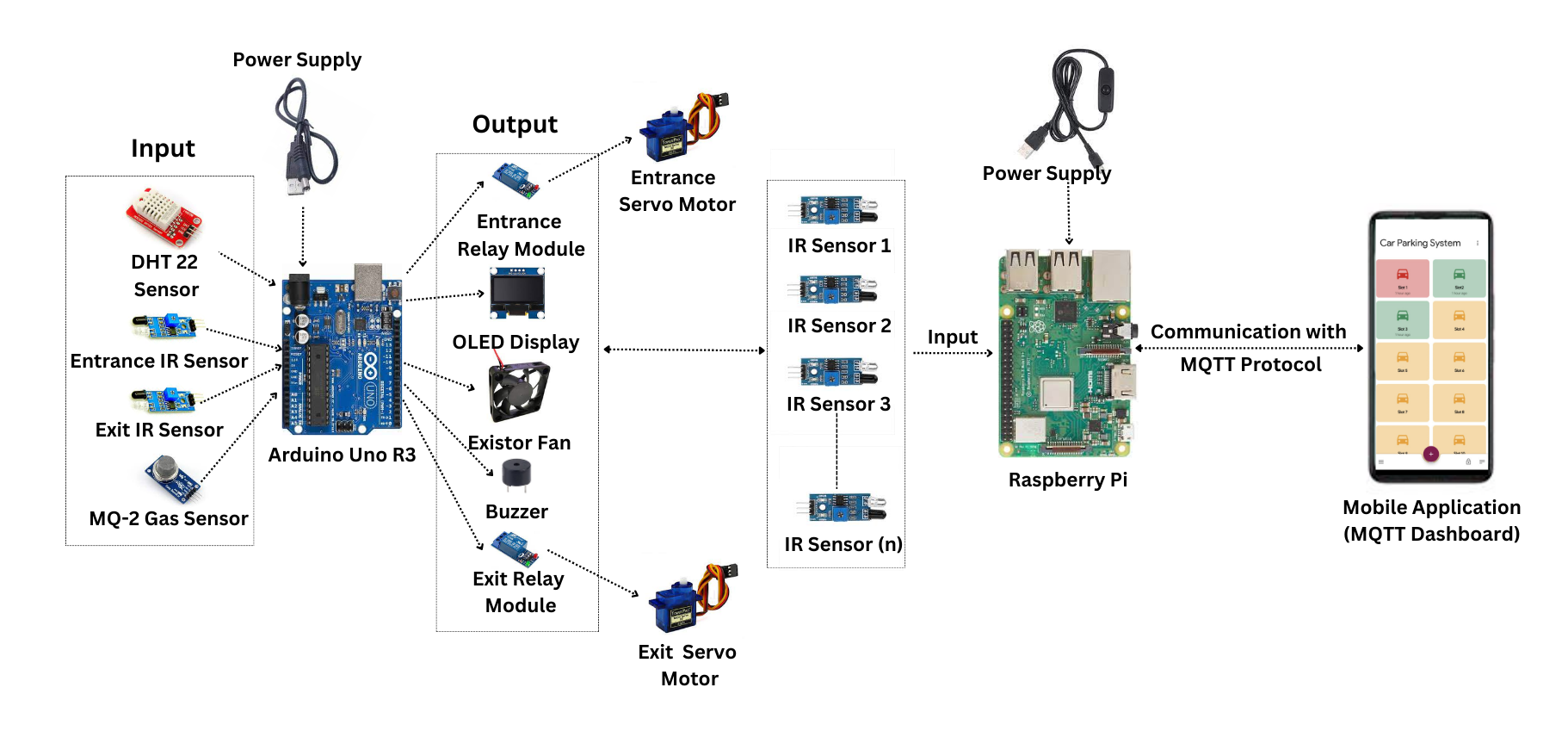}}
\caption{Proposed Block Diagram for Controlling each Connection}
\label{fig:f1}
\end{figure*}

In our system, we implement an advanced IoT-enabled solution integrating various sensors and MCUs for advanced parking management. Each slot has an IR sensor for vehicle detection, ensuring accurate entry and exit monitoring. A DHT22 sensor continuously measures temperature and humidity, displaying data on an OLED screen. An MQ2 gas sensor detects hazardous gases, activating an exhaust fan when needed. MQTT protocol facilitates real-time updates on slot availability and environmental conditions via a mobile app \cite{arefin2023iot}. These components' comprehensive workflow and interactions are depicted in Fig. \ref{fig:f1}, illustrating the block diagram of our IoT-based smart car parking system.

\subsection{Collecting of Hardware Requirements}
This system requires several essential hardware components to function effectively. Table \ref{tab:t2} summarizes the specific roles and functionalities of each hardware component within the smart car parking system.

\begin{table}[H]
\centering
\scriptsize 
\tiny
\caption{Functionalities of Hardware Components in the System}
\label{tab:t2}
\begin{tabular}{|l|p{6cm}|}
\hline
\textbf{Component} & \textbf{Functionality} \\
\hline
Arduino Uno & Controls entrance and exit gates with IR sensors. Manages servo motors and integrates MQ2 gas sensors for hazardous gas monitoring. Activates exhaust fan and buzzer for safety alerts. \\
\hline
IR Sensors & Detects vehicles at entrance and exit gates. Monitors occupancy status of parking slots. \\
\hline
Relay Module & Interfaces with Arduino Uno to control servo motors for gate operation. \\
\hline
Servo Motors & Operates entrance and exit gates based on signals from Arduino Uno via a relay module. \\
\hline
MQ2 Gas Sensor & Monitors gas concentrations in the parking area. Triggers exhaust fan via Arduino Uno for ventilation. \\
\hline
Exhaust Fan & Activates based on MQ2 sensor readings for proper ventilation. \\
\hline
Buzzer & Alerts audibly when vehicles approach the entrance gate. Connected to Arduino Uno. \\
\hline
DHT22 Sensor & Measures temperature and humidity inside the parking facility. Provides real-time data to Arduino Uno. \\
\hline
OLED Display & Shows real-time environmental data from the DHT22 sensor. Displays parking slot availability updates. \\
\hline
Raspberry Pi & Collects data from IR sensors at parking slots. Sends real-time updates via MQTT to the mobile app. Enables monitoring and management of parking slots. \\
\hline
\end{tabular}
\end{table}

\subsection{Arrangement of Software Requirements}
\subsubsection{Arduino IDE}
The Arduino IDE supports Linux, Mac OS X, and Windows, facilitating programming in C and C++. Programs, known as sketches, are created and uploaded to Arduino boards via the IDE for execution \cite{arduino_ide_v1_tutorial}.

\subsubsection{Raspberry Pi OS}
Raspberry Pi OS is a Debian-based Unix-like OS for Raspberry Pi boards (excluding Pico). It features a modified LXDE desktop with Openbox and comes with Wolfram Mathematica, VLC, and a lightweight Chromium browser \cite{islam2024iot}.

\subsubsection{Putty and Advanced IP scanner}
To connect and upload code to a Raspberry Pi (RPi), we use tools like PuTTY and Advanced IP Scanner. First, we find the RPi's IP address with the Advanced IP Scanner. Then, we open PuTTY, enter the IP to SSH in, and log in. Navigate to our code directory using terminal commands (cd) and use PuTTY's SCP feature to transfer files from our local machine to the RPi \cite{iotstarters-putty-rpi, rahman2020distb}.

\subsubsection{Mobile Application}
MQTT Dashboard is an Android application designed for easy interaction, control, and management of MQTT-enabled devices. The app offers customizable interface components, referred to as tiles, allowing users to create personalized MQTT dashboards tailored to their needs \cite{mqtt-dashboard-docs}.

\subsection{Analysis of System Design}
Our IoT-enabled Parking System efficiently manages space availability, monitors environmental conditions, and ensures proper ventilation, enhancing user experience and maintaining optimal conditions. It uses a Raspberry Pi for MQTT communication and an Arduino Uno for controls and sensor integrations. In Algorithm \ref{alg:pseu}, The pseudocode represents the logic for the smart car parking system. It outlines how the system manages parking slots, updates displays, and communicates status through MQTT. 

\begin{algorithm}[h]
\scriptsize
\caption{IoT-Enabled Smart Car Parking System}
\label{alg:pseu}
\begin{algorithmic}[1]
    \State Init(S, T, H, G, F, D) \Comment{Initialize components}
    \State $S = [0, 0, ..., 0]$ \Comment{Set all parking slots to vacant}
    \While {Active}
        \If {IR\_in == 1} \Comment{Car detected at entrance}
            \State $Total\_vacant -= 1$ \Comment{Decrease vacant slots}
            \If {$Total\_vacant \geq 0$} \Comment{Slots available}
                \State $G\_in = 1$ \Comment{Open entrance gate}
                \State $B = 1$ \Comment{Turn on buzzer}
                \State $D(T, H, Total\_vacant, Total\_slots)$ \Comment{Update Display}
                \State $MQTT(T\_P, S)$ \Comment{Publish status to MQTT}
            \Else \Comment{No slots available}
                \State $Total\_vacant += 1$ \Comment{Revert vacant slots}
            \EndIf
        \EndIf
        \If {IR\_out == 1} \Comment{Car detected at exit}
            \State $G\_out = 1$ \Comment{Open exit gate}
            \State $Total\_vacant += 1$ \Comment{Increase vacant slots}
            \State $D(T, H, Total\_vacant, Total\_slots)$ \Comment{Update Display}
            \State $MQTT(T\_P, S)$ \Comment{Publish status to MQTT}
        \EndIf
        \For {$i = 1$ to $n$} \Comment{Update slot status}
            \State $S_i = IR\_slot_i$ \Comment{Read slot status}
            \State $MQTT(T\_P_i, S_i)$ \Comment{Publish slot status}
        \EndFor
        \State $T, H = DHT22$ \Comment{Get temperature and humidity}
        \State $D(T, H, Total\_vacant, Total\_slots)$ 
        \State $G = MQ2$ \Comment{Read gas level}
        \If {$G > G\_th$} \Comment{Gas threshold exceeded}
            \State $F = 1$ \Comment{Activate exhaust fan}
        \EndIf
    \EndWhile
    \State Deinit(S, T, H, G, F, D) \Comment{Shutdown components}
\end{algorithmic}
\end{algorithm}

\begin{figure*}[!htb]
\centerline{\includegraphics[scale=0.23]{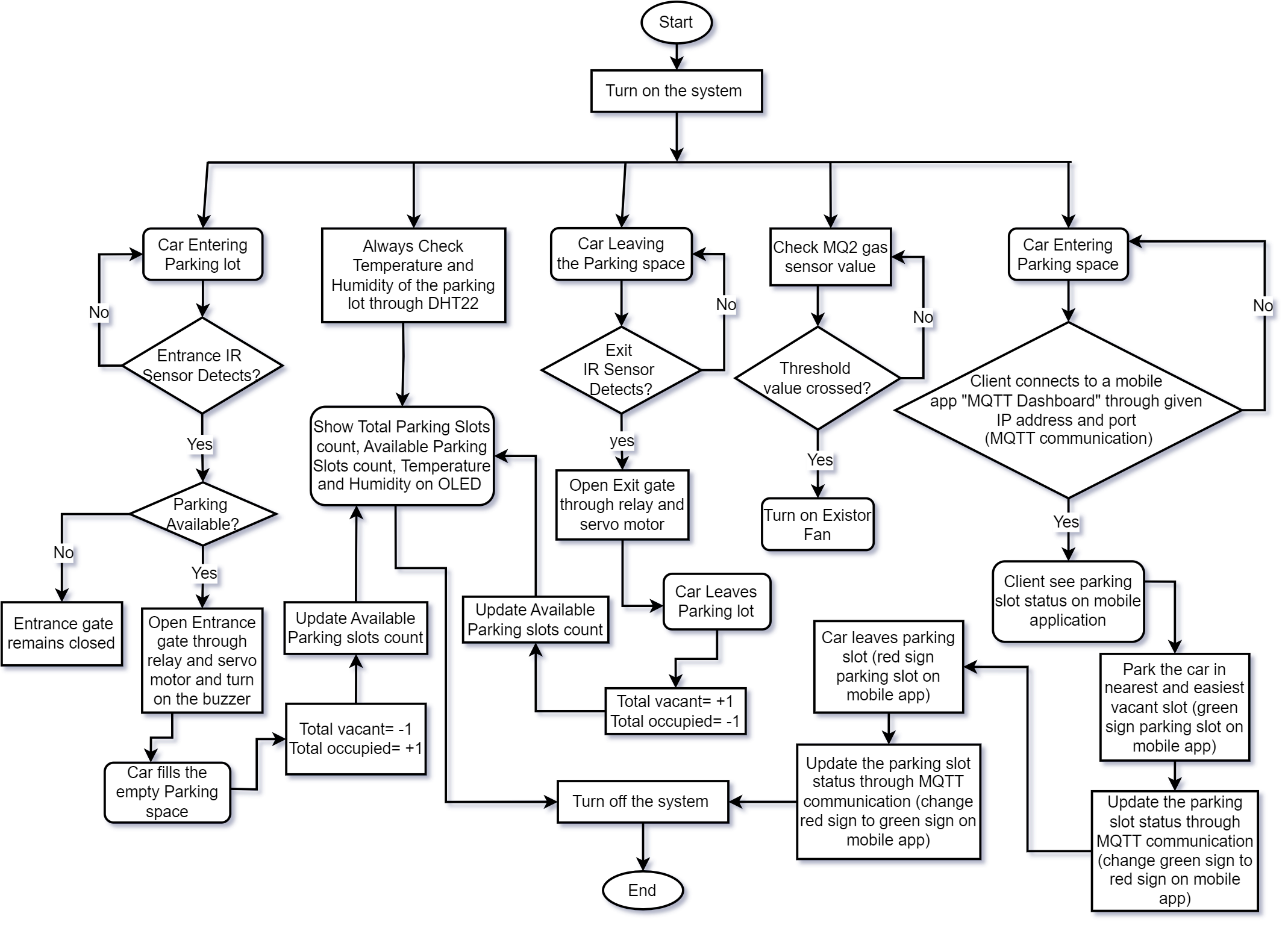}}
\caption{Workflow Diagram of Proposed System}
\label{fig:f2}
\end{figure*} 

When activated, the system begins monitoring all sensors and controls. As a car approaches, the entrance IR sensor detects the vehicle ($V_{\text{in}}$). If no car is detected ($V_{\text{in}} = 0$), the entrance gate remains closed. If a car is detected ($V_{\text{in}} = 1$), the system checks for available parking spaces. If no space is available ($\text{Total vacant} = 0$), the entrance gate remains closed. If a space is available ($\text{Total vacant} > 0$), the entrance gate opens via a relay and servo motor, and a buzzer activates ($B = 1$). 
The system continuously monitors temperature ($T$) and humidity ($H$) using a DHT22 sensor. This information, along with the parking slots count, is displayed on an OLED screen (Equation \ref{eq2}):

\begin{equation}
\text{Display} = f(T, H, \text{Total vacant}, \text{Total slots})
\label{eq2}
\end{equation}

When a car exits, the exit IR sensor detects the vehicle ($V_{\text{out}}$). If no car is detected ($V_{\text{out}} = 0$), the exit gate remains closed. If a car is detected ($V_{\text{out}} = 1$), the exit gate opens via a relay and servo motor. The available slots count is then updated.

The system integrates with a mobile app \enquote{MQTT Dashboard} via MQTT communication. Clients connect to the app using the provided IP address and port ($\text{IP}_{\text{app}}, \text{port}_{\text{app}}$). The real-time status of parking slots is communicated as follows:

The MQTT communication between the Raspberry Pi and the mobile app involves publishing and subscribing to topics. Let:
\begin{itemize}
    \item $S$ be the state of the parking slot ($S = 1$ for occupied, $S = 0$ for vacant).
    \item $T_P$ be the topic for parking slot status updates.
    \item $D_R$ be the data rate, $\text{delay}$ be the communication latency, and $EC$ be the error correction rate.
\end{itemize}

The parking slot status is updated and published to the topic $T_P$ (Equation \ref{eq4}):

\begin{equation}
\text{Publish}(T_P, S) \implies \text{Message}(S)
\label{eq4}
\end{equation}

Clients subscribe to the topic to receive updates (Equation \ref{eq5}):

\begin{equation}
\text{Subscribe}(T_P) \implies \text{Receive Message}(S)
\label{eq5}
\end{equation}

The data transfer rate ($D_R$) for MQTT communication is given by (Equation \ref{eq6}):

\begin{equation}
D_R = \frac{\text{Data transmitted}}{\text{Time}}
\label{eq6}
\end{equation}

The communication delay ($\text{delay}$) is calculated as (Equation \ref{eq7}):

\begin{equation}
\text{delay} = \text{time}_{\text{end}} - \text{time}_{\text{start}}
\label{eq7}
\end{equation}

The error correction rate ($EC$) is given by (Equation \ref{eq8}):

\begin{equation}
EC = \frac{\text{Errors corrected}}{\text{Total errors}}
\label{eq8}
\end{equation}

Upon entering the parking lot, clients can view available slots on their mobile devices and park in the nearest and easiest vacant slot. The parking slot status is updated through MQTT communication, changing from green (available) to red (occupied) on the mobile app. Additionally, the system includes an MQ2 gas sensor to monitor the parking lot for hazardous gas levels ($G$). If the gas concentration exceeds a predefined threshold ($G > G_{\text{threshold}}$), an exhaust fan activates ($F = 1$) to ventilate the area (Equation \ref{eq9}):

\begin{equation}
G > G_{\text{threshold}} \implies F = 1
\label{eq9}
\end{equation}

The probability of the parking lot being fully occupied ($P_{\text{full}}$) can be modeled using a Poisson distribution with arrival rate $\lambda$. The probability that the number of cars $k$ equals the total slots $n$ is (Equation \ref{eq10}):

\begin{equation}
P_{\text{full}} = \frac{\lambda^n e^{-\lambda}}{n!}
\label{eq10}
\end{equation}

If the average time a car spends in the lot is $T_{\text{avg}}$, and the arrival rate is $\lambda$, then the average number of cars in the lot ($L$) can be estimated using Little’s Law (Equation \ref{eq11}):

\begin{equation}
L = \lambda \cdot T_{\text{avg}}
\label{eq11}
\end{equation}

The response time of the ventilation system ($t_{\text{response}}$) when gas concentration exceeds the threshold is (Equation \ref{eq12}):

\begin{equation}
t_{\text{response}} = \frac{\Delta G}{r}
\label{eq12}
\end{equation}

Where $\Delta G$ is the difference between the current gas concentration and the threshold, and $r$ is the rate of gas concentration reduction. When turned off, all monitoring and control operations cease, concluding the workflow.

\subsection{Working Procedure of System Model}
Our IoT-enabled Smart Car Parking System is designed to efficiently manage parking availability, monitor environmental conditions, and provide real-time updates to users through a mobile application. The flowchart in Fig. \ref{fig:f2} illustrates the whole System. This kind of system allows data handling to be more flexible and real-time and enhances user experience in maintaining ideal conditions within the parking lot as well.\vspace{2mm}

When the system is turned on, it starts checking both temperature and humidity in the parking lot using a DHT22 sensor (it also has an OLED screen that will show data of availability); It has an entrance IR sensor that detects vehicles coming in. When a car is detected and there is an available parking space, the entrance gate opens with a relay/servo motor on hand while another buzzer goes off. The slot count changes to the available slots. The system uses MQTT communication with IR sensors and integrations with \enquote{MQTT Dashboard} mobile app. Using the IP and port to connect with this module, users can view parking slot status online. Upon arrival, they will note which slots are open and park there. The slot status on the app changes from green (free) to red (occupied). Exit IR sensor detects a car exiting, the gate opens, and the slot count decreases. The app status changes from red to green, indicating that the slot is now available. The system also comprises a gas sensor MQ2, which will detect the level of Hazardous Gas in the Parking slot. When the gas concentration is high over a preset point of reference, an exhaust fan triggers and forcibly evacuates exhaust, allowing all personnel to work safely in that space. 
Finally, the system can be turned off, ending all monitoring and control operations, thus concluding the workflow of the smart parking system. This structured and automated approach enhances efficiency, safety, and user satisfaction within the parking loT.


\section{Result Analysis and Performance Measurement}
\subsection{Performance Analysis}
\subsubsection{Performance Analysis of IR Sensor}
\begin{figure}[h]
\centerline{\includegraphics[scale=0.09]{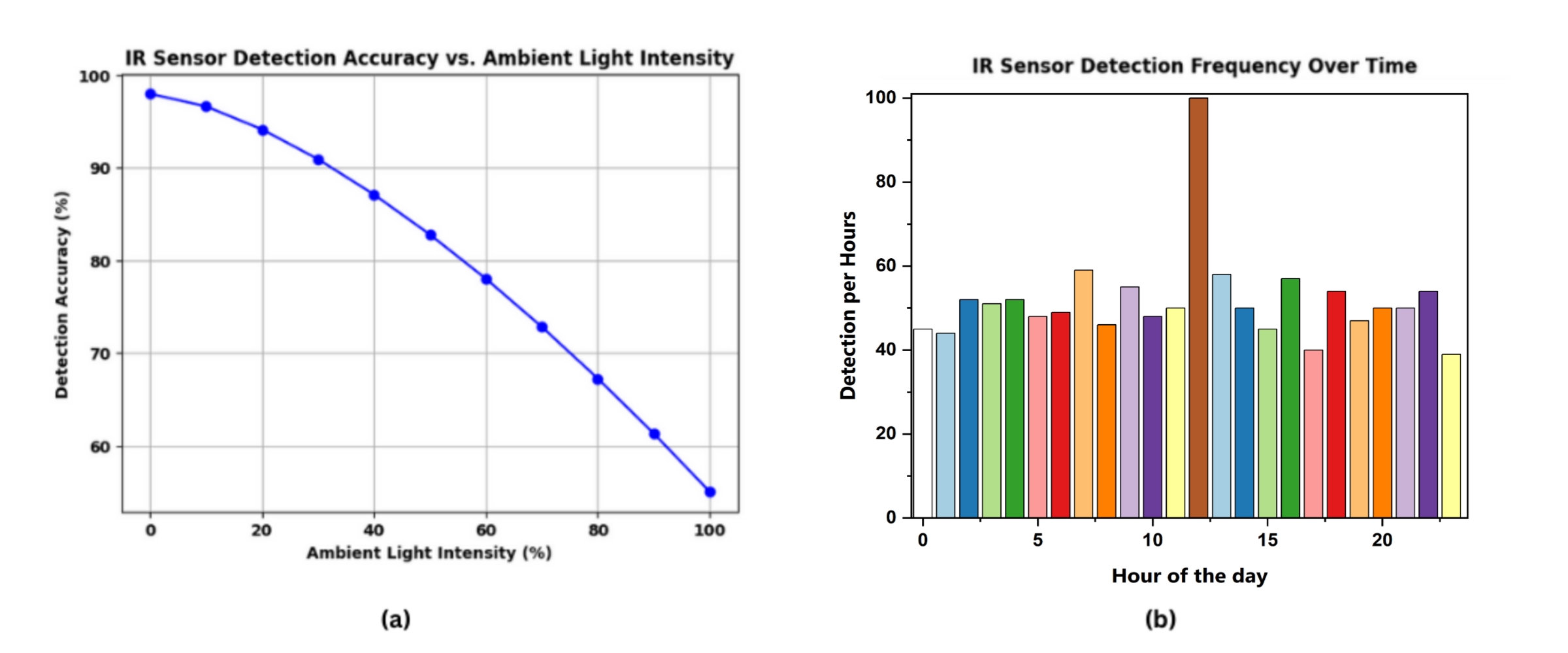}}
\caption{(a) IR Sensor Detection Accuracy vs. Ambient Light Intensity, (b) IR Sensor Detection Frequency}
\label{fig:f3}
\end{figure}

Fig. \ref{fig:f3}(a) analyzes how an IR sensor can detect vehicle entry and exit with different light intensities for a whole week. The sensor’s accuracy reduced by a whopping 52\% to 98\% of what it initially was as the ambient light intensity got higher and, therefore, the calibration and screening became necessary, which is why the sensor’s performance exhibited. For optimal parking management, it is important to provide high detection accuracy, which will reduce false detection.

\vspace{2mm}
Fig. \ref{fig:f3}(b) shows the frequency of IR sensor detection in the CMH parking area, Dhaka Cantonment, Bangladesh
over 24 hours. The peak detection is up to 100 per hour between 12 PM and 1 PM. It is used to help manage the flow of traffic, allocate resources, maintain system efficiency, etc. Knowing peak hours helps to optimize the schedules and allow for dynamic pricing.

\subsubsection{Performance Analysis of DHT22 Temperature and Humidity Sensor}
\begin{figure}[h]
\centerline{\includegraphics[scale=0.25]{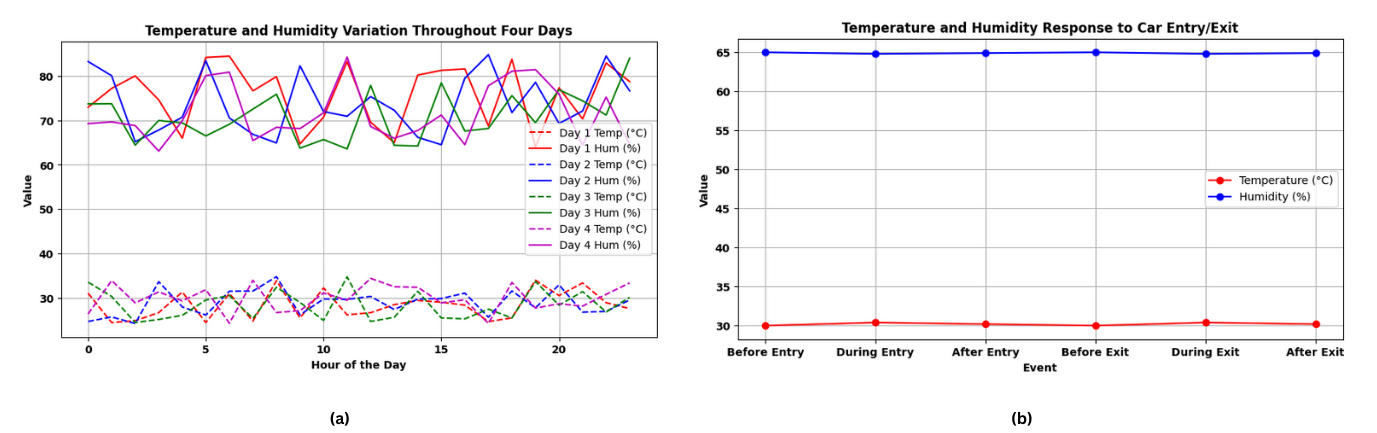}}
\caption{(a) Temperature and Humidity Value Monitoring, (b) Temperature and Humidity Response to Car Entry/Exit}
\label{fig:f4}
\end{figure}

Fig. \ref{fig:f4}(a) shows the temperature and humidity variation in the parking area for four days with temperature ranging from 24°C to 35°C, Humidity was varied between 63\% and 85\%. The entrance and exit of vehicles, each step also introduces temperature humidity until the degree of comfort as well as safety used in car parking areas.

On the other hand, Fig. \ref{fig:f4}(b) illustrates temperature and humidity measurements before car entry and exit distinguished into the different states. Upon entry, the temperature rose by 0.4°C and the humidity by 0.2\%, with a slight decrease of both upon leaving. By getting insights into those alterations, appropriate conditions for the facilities can be established.

\vspace{2mm}

\subsubsection{Performance Analysis of MQ-2 Gas Sensor}
\begin{figure}[h]
\centerline{\includegraphics[scale=0.10]{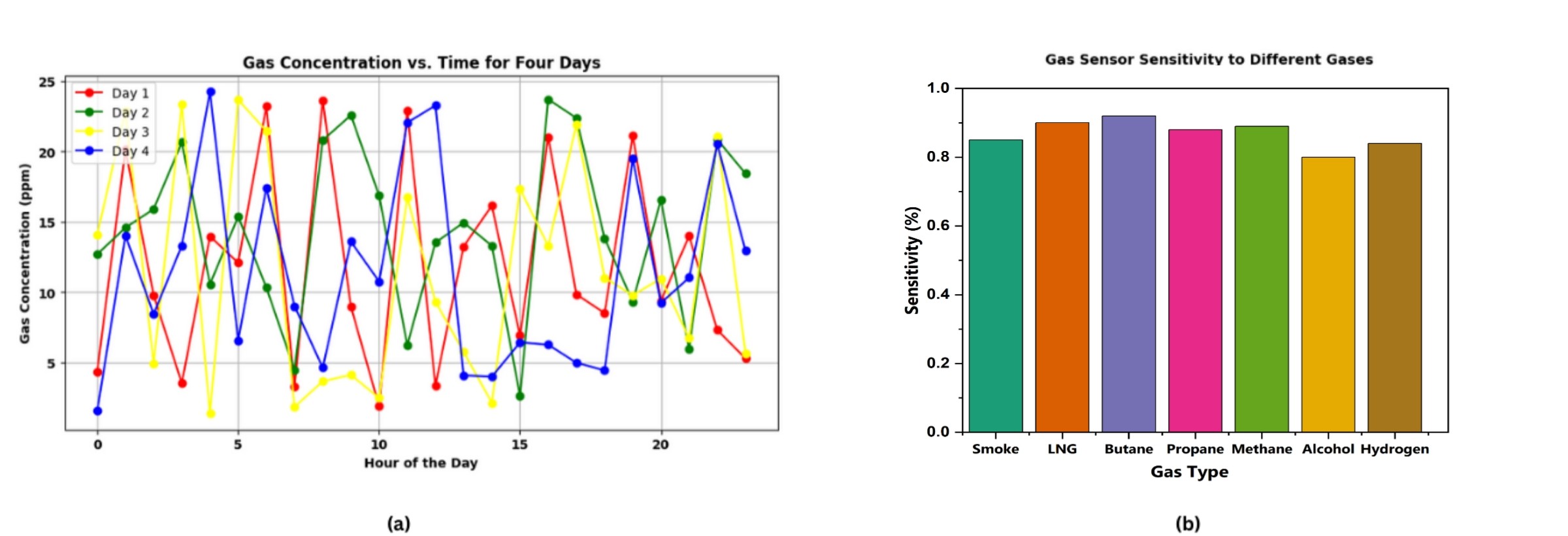}}
\caption{(a) Gas Value Monitoring, (b) Comparison of Different Gas Sensor's Sensitivity}
\label{fig:f5}
\end{figure}

Fig. \ref{fig:f5}(a) shows the gas concentration levels (e.g., CO) in the parking area for four days ranging from 1 ppm to 25 ppm at peak hours. This information is very critical to maintain air quality and safety. 

Fig. \ref{fig:f5}(b) is a graph on the sensitivity of the MQ-2 sensor in relation to various gases, where the highest sensitivity is to Butane — 0.92 and the lowest to Alcohol — 0.80. This ensures comprehensive air quality monitoring.

\subsection{Comparison with IoT and Non-IoT System}
\begin{figure}[h]
\centerline{\includegraphics[scale=0.28]{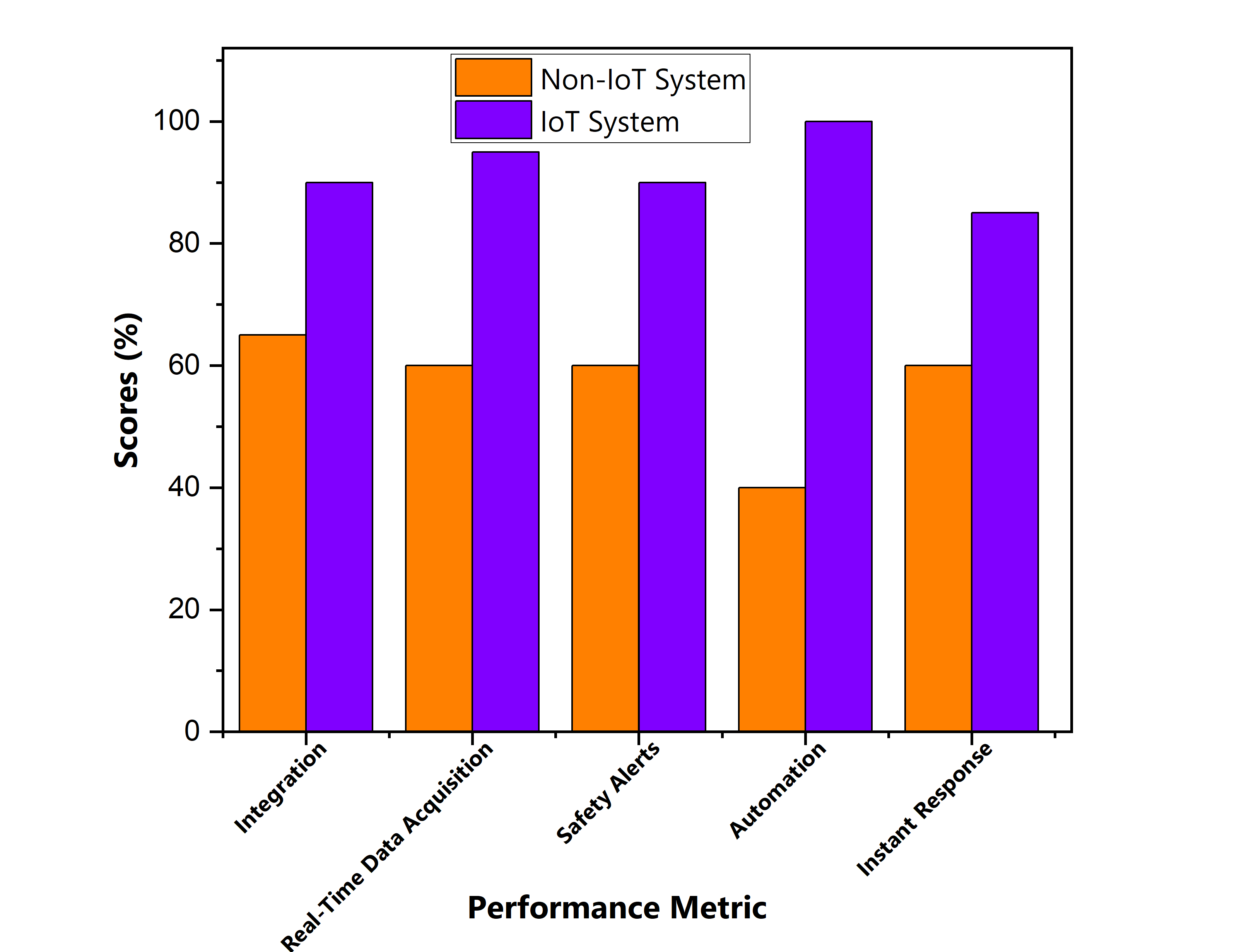}}
\caption{Comparison the proposed system with Non-IoT System}
\label{fig:f15}
\end{figure}

Fig. \ref{fig:f15} shows the difference in the efficiency of a car parking system, which is the IoT one and the non-IoT one, and also shows the problems. The IoT-enabled parking system is advanced in connectivity, immediate data acquisition, alert security, automation, and quick response, thus improving efficiency and the user experience. Table \ref{table:comparison} details the performance metrics, illustrating the significant improvements achieved by the IoT system \cite{rahman2023impacts}.

\begin{table}[h]
\centering
\scriptsize
\caption{Comparison between IoT and Non-IoT Systems}
\begin{tabular}{|l|c|c|}
\hline
\textbf{Performance Metric}         & \textbf{Non-IoT System(\%)} & \textbf{IoT System(\%)} \\ \hline
Integration                         & 65                & 90            \\ \hline
Real-Time Data Acquisition          & 60                & 95            \\ \hline
Safety Alerts                       & 60                & 90            \\ \hline
Automation                          & 40                & 100           \\ \hline
Instant Response                    & 60                & 85            \\ \hline
\end{tabular}
\label{table:comparison}
\end{table}

\subsection{Real-time Analysis}
This smart system's real-time analysis requires constant monitoring and reporting of various parameters that are necessary for the perfect functioning of the parking facility. 

\subsubsection{Mobile Application Integration}
\begin{figure}[h]
\centerline{\includegraphics[scale=0.18]{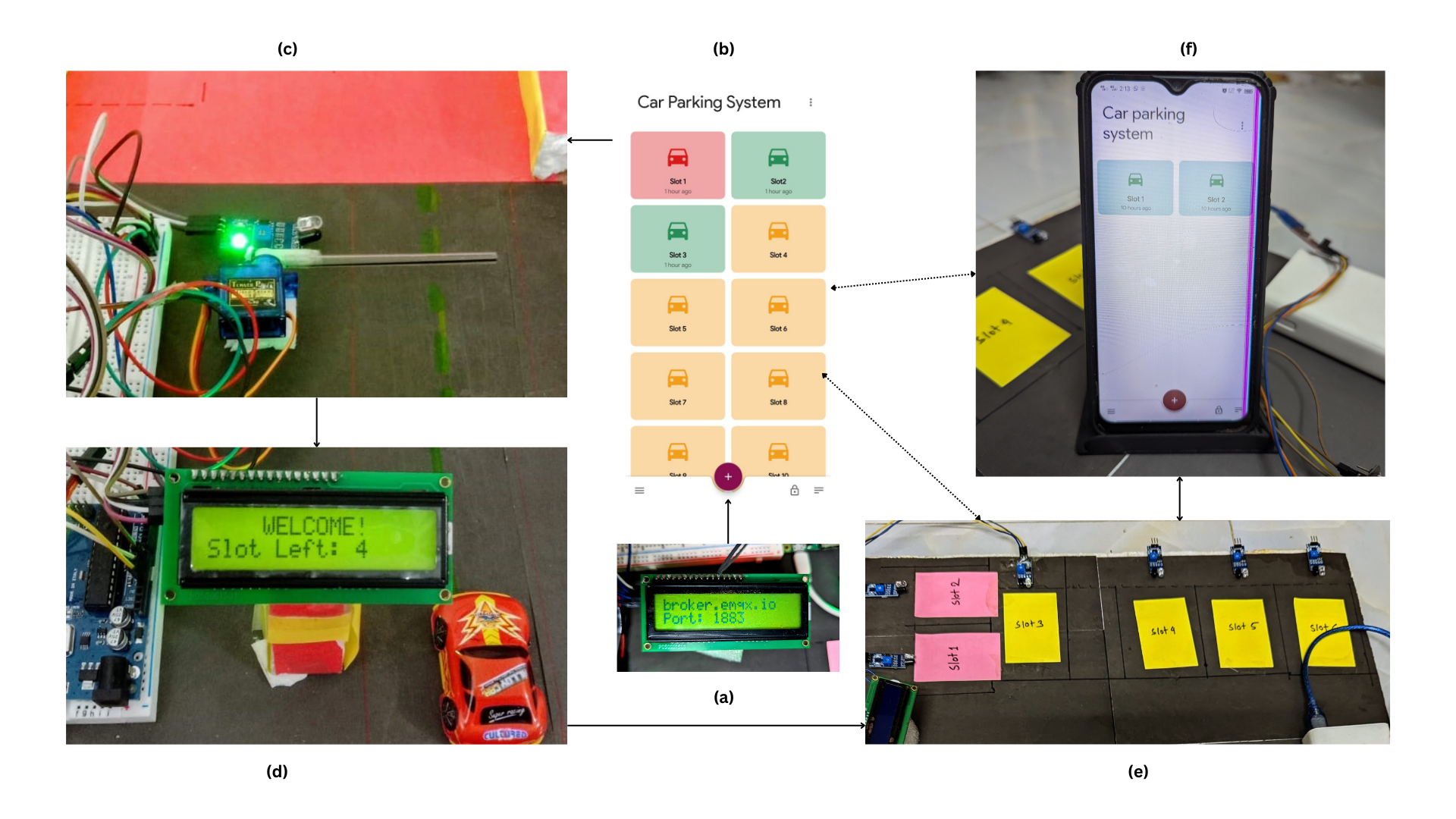}}
\caption{(a) Displayed IP Address and Port Number, (b) Mobile Application (MQTT Dashboard), (c) Exit Gate of Parking Area, (d) Displayed Number of Free Available Slots, (e) Slots Equipped with IR Sensor, (f) Send Sensor Data to Mobile App}
\label{fig:f16}
\end{figure}

Drivers connect to the mobile application using a specific port number and IP address, enabling them to easily locate the nearest vacant parking space. The port number and IP address are displayed in Fig. \ref{fig:f16}(a). Sensor data is transmitted using the MQTT protocol to a mobile application, which drivers can use to view the status of each parking slot. Fig. \ref{fig:f16}(b) shows the application MQTT Dashboard displays vacant slots in green and occupied slots in red. Other colors indicate the inactive slot. 

\subsubsection{Entrance and Exit Gate Management} When a car comes towards the entrance, an IR-sensor detects that there is a vehicle. A buzzer goes off, notifying the car park attendant that a vehicle is wanting access to said area. Fig. Fig.\ref{fig:f16}(c) illustrates the implementation of a servo motor acting as an exit gate using an IR sensor \cite{rahman2023towards}. At the entrance, there is an OLED display with how many slots exist in total and available for a car to park before it, which dynamically updates upon coming vehicles. Fig. \ref{fig:f16}(d) shows the number of all slots offered by the parking area.

\subsubsection{Parking Slot Monitoring}
Each parking slot is equipped with an IR sensor to detect the presence or absence of a vehicle. Fig. \ref{fig:f16}(e) shows the implementation of each slot with an IR sensor. Fig. \ref{fig:f16}(f) shows the sensors continuously send status updates to the central system and Mobile Application (MQTT Dashboard).

\subsubsection{Environmental Monitoring}
Real-time temperature and humidity data are generated using a DHT22 sensor, which continuously measures the environment and displays them on an OLED panel. The concentration of gas is measured by the MQ-2 gas sensor. This device will activate in order to release part of that concentrated air through the exhaust fan in the event that the gas levels surpass a certain threshold, preserving safety.

\section{Discussion, Limitations and Future Work}
Our IoT-enabled car parking system significantly improves performance over traditional systems. By integrating IR sensors, servo motors, an OLED display, and a mobile app with environmental sensors (DHT22, MQ-2 Gas Sensor), it creates an efficient parking management solution. Real-time data acquisition and seamless communication enhance efficiency, while automation improves user experience and safety. Drivers receive instant updates on parking availability, and the system maintains optimal environmental conditions. 


Our system indeed encounters some certain sensor-related issues. IR sensors can lose precision because of dust and light, which make them inaccurate and, therefore, need to be calibrated frequently. The DHT22 sensor as well shifts with time, distancing equally its temperature and humidity reading abilities and will require some time adjustment. The MQ-2 gas sensor deteriorates with small abuse, and its efficiency is dispersed if it is not replaced periodically. According to the feedback of the users, it can be observed that the system also relies so much on the network, such that even a small blackout leads to the suspension of real-time data, hence lowering efficiency.


In order to tackle these limitations, we intend to utilize IR sensors that glory in self-cleaning design devoid of manual intervention due to dust and lighting alterations. In the case of a DHT-22 sensor, we suggest safe methods of changing the sensor age barriers, none of which will improve the calibration of the sensor to inferior levels. In view of gas sensors, backward and nerdy redundant MQ-2 will be added if one of the gas detectors turns out to be faulty. To solve the issue of consistency in network reliability, we’ll allow an option of offline storage of data when any of the network connectivity is not in effect. Furthermore, the system will also include predictive maintenance of sensors using machine learning, security, and safeguarding of the system’s data and transmission with end-to-end encryption as well as using blockchain technology.

\section{Conclusion}
Our IoT-enabled car parking system is capable of functioning better than traditional non-IoT Systems that make use of IR sensors, Servo Motors, OLED Displays, DHT22 sensors, MQ-2 gas sensors, and mobile apps. They fetch data in real-time communication with each other and control automatically, leading to an improvement in productivity, which is why it mitigates user experience and requires security. With initial setup costs and network dependence now, future work will be aimed at reducing the cost of this system as well as enhancing its reliability and security. Through the system, smart city initiatives will be boosted since they reduce emissions and support urban mobility by optimizing the use of parking spaces. This demonstrates the transformative power of IoT as part of urban infrastructure, with possible future applications ranging from traffic management to energy and public safety solutions that improve the quality of life in cities globally.

\bibliographystyle{IEEEtran}

\bibliography{sample}

\end{document}